\documentclass[aps,preprint,prd,showpacs,nofootinbib]{revtex4}
\usepackage{amsmath}
\usepackage{graphicx}
\usepackage{dcolumn}
\usepackage{bm}
\usepackage{amssymb}
\usepackage{latexsym}
\usepackage{color}

\def\be{\begin{equation}}
\def\ee{\end{equation}}
\def\ba{\begin{eqnarray}}
\def\ea{\end{eqnarray}}

\begin{document}

\title{Parity violation in pre-inflationary bounce    }

\author{Yu-Tong Wang$^{1}$\footnote{Email: wangyutong12@mails.ucas.ac.cn}}
\author{Yun-Song Piao$^{1,2}$\footnote{Email: yspiao@ucas.ac.cn}}

\affiliation{$^1$ School of Physics, University of Chinese Academy
of Sciences, Beijing 100049, P.R.China}
\affiliation{$^2$ State Key Laboratory of Theoretical Physics, Institute of Theoretical Physics, \\
Chinese Academy of Sciences, P.O. Box 2735, Beijing 100190, China}

\begin{abstract}

The power suppression on large scale in the CMB TT-mode power
spectrum might imply the occurrence of a pre-inflationary bounce.
We calculate the circularly polarized gravitational wave, leaded
by the gravitational Chern-Simons term universally appearing in
particle physics and string theory, in the inflation model with
the pre-inflationary bounce. The circularly polarized
gravitational wave will induce TB and EB-mode correlations at CMB
last scatting surface. We find that if the pre-inflationary bounce
actually occurs, the TB and EB-mode correlations on large scale
will be enhanced, while the BB-mode correlation on corresponding
scales is suppressed.

\end{abstract}

\maketitle

\section{Introduction}

Recently, the Planck collaboration has reported a power deficit in
the CMB TT power spectrum on largest scale
\cite{Ade:2013uln},\cite{Ade:2013nlj}, which also was found in
WMAP data, and is not concordant with the Planck bestfit model.
Its statistical significance is about $ 3\sigma$. In addition, the
Planck collaboration has also reported a hemispherical power
asymmetry in CMB at low-$l$ \cite{Ade:2013nlj}, which conformed a
similar obsevation of WMAP \cite{Eriksen:2007pc,Hoftuft:2009rq},
see also \cite{Rassat:2014yna},\cite{Akrami:2014eta}.

These large-scale anomalies might be a hint of the
pre-inflationary physics relevant with the initial singularity
\cite{Piao:2003zm},\cite{Dudas:2012vv}. In
Refs.\cite{Piao:2003zm},\cite{Falciano:2008gt},\cite{Mielczarek:2008pf},\cite{Liu:2013kea},\cite{Qiu:2014nla},
the pre-inflationary universe is in a contracting phase and after
the bounce the universe begins to inflate.
In
Refs.\cite{Liu:2013iha},\cite{Biswas:2013dry},\cite{Labrana:2013oca},
the pre-inflationary universe is in a superinflationary phase, see
also \cite{Liu:2014tda} for the case with $\epsilon\ll -1$, which
is similar to the emergent universe scenario \cite{Ellis:2002we}.
Above pre-inflationary evolutions generally will generate a
large-scale cutoff in the primordial power spectrum, which may
naturally suppress the CMB TT power spectrum at low-$l$, see e.g.
\cite{Liu:2013kea},\cite{Liu:2013iha} for the details. In such
scenarios, it is generally required that the slow-roll inflation
lasts for just the minimal number of efoldings, i.e.just enough
inflation \cite{Ramirez}, and thus the power deficit at low-$l$
may be attributed to the evolution of the pre-inflationary
non-slow-roll background, e.g. see Ref.\cite{Cicoli:2014bja} for a
discussion on the pre-inflationary expanding phase, and also
earlier Refs.\cite{Contaldi:2003zv} and \cite{Cline:2003ve}. See
e.g.
\cite{Liu:2013iha},\cite{Kitazawa:2014dya},\cite{Cicoli:2013oba},\cite{Pedro:2013pba},\cite{Piao:2003hh}
for some stringy embeddings.

The measure of large-scale E and B-mode polarizations
\cite{Seljak:1996gy},\cite{Kamionkowski:1996zd} will help to
provide a unambiguous test for the pre-inflationary evolution.
The scalar perturbation contributes the TT, TE and EE-mode
correlations in CMB. While the gravitational wave (GW)
contributes the BB-mode correlation besides the TT, TE and EE-mode correlations.




However, if the gravity is chiral, we might have other channels to
test the evolution of pre-inflationary universe. The gravitational
Chern-Simons (gCS) term, motivated by the anomaly cancelation in
particle physics and string theory
\cite{Witten83},\cite{Jackiw:2003pm}, is parity-violating, see
Ref.\cite{Alexander:2009tp} for a review, which will produce a
difference between the amplitudes of right-handed and left-handed
GWs \cite{Lue:1998mq},\cite{Choi:1999zy},\cite{Alexander:2004us}.
The primordial circularly polarized GW will induce TB and EB-mode
correlations at CMB last scatting surface
\cite{Saito:2007kt},\cite{Li:2009rt},\cite{Gluscevic:2010},\cite{Xia:2012},\cite{AQWT}.

The TB and EB correlations may be also brought by the
electromagnetic CS term
e.g.\cite{Feng:2006dp},\cite{Liu:2006uh},\cite{Cabella:2007br},
which affects the CMB polarizations after the photon decoupling
e.g.\cite{Li:2008tma}, see \cite{Li:2014oia} for the analysis with
latest data. As a result, the shape of TB-mode power spectrum is
generally the same with that of TE-mode power spectrum. Thus such
a power spectrum lead by electromagnetic CS term is different from
a power spectrum lead by gravitational CS term, see e.g.\cite{Gluscevic:2010}.

In conventional slow-roll inflation scenario, the circular
polarization of primordial GW from the gCS term is negligible
\cite{Alexander:2004wk},\cite{Lyth:2005jf}. Thus the primordial TB
and EB correlations in CMB is unseen
\cite{Saito:2007kt},\cite{Gluscevic:2010}. However, the
significant circular polarization may be created in a
string-inspired inflationary model with the GB term
\cite{Satoh:2007gn}. In this sense, it seems that the TB and
EB-mode correlations recording the chirality of primordial gravity
might also encode the information of the evolution of primordial
universe. The pre-inflationary bounce
\cite{Piao:2003zm},\cite{Liu:2013kea} not only may account for the
CMB anomalies at large scale, but also avoid the initial
singularity problem of inflationary universe. Thus it is
interesting to investigate the TB and EB-mode correlations in such
a scenario.

Here, we will calculate the circularly polarized GW leaded by the
gCS term in the inflation model with the pre-inflationary bounce.
We find that if the pre-inflationary bounce actually occurs, the
TB and EB-mode correlations on large scale will be enhanced, while
the BB-mode correlation on corresponding scales is suppressed.



\section{Pre-inflationary gravitational wave  }

The gravitation action including the gCS term is \be S={
S}_{Einstein}+\int (-g)^{1/2}d^4x {f( \phi) \over 8} R\wedge R,
\ee in which $\phi$ is identified as the inflaton in the slow-roll
inflation or the background field in the pre-inflationary
evolution.

The gCS term only affects the tensor perturbation, but does not
affect the scalar perturbation and the evolution of background,
e.g.\cite{Lue:1998mq}. The tensor perturbation $h_{ij}$ obeys
$\delta^{ij}h_{ij}=0$ and $\partial_i h^{ij}=0$, and its action is
\ba S_2 & = & {1\over 8}\int d\eta d^3x \bigg[ a^2 M_P^2
\left({{h_{ij}}^\prime}^2-(\partial
{h_{ij}})^2\right) - 
f^\prime \epsilon^{ijk}\left({h^q_{i}}^\prime (\partial_j
h_{kq})^\prime-\partial^r h^{q}_{i}\,\partial_{j}\partial_r
h_{kq}\right)\bigg] , \label{action1}\ea where $\epsilon^{ijk}$ is
the Levi-Cevita symbol, and $'$ is the derivative with respect to
$\eta=\int dt/a$.

Here, the universe is initially in a contracting phase and after
the bounce it is in the inflationary phase. To investigate the
evolution of $h_{ij}$,
we will adopt an instantaneous matching between both phases
\cite{Piao:2003zm},\cite{Liu:2013kea}, i.e.
\ba a & \simeq & a_*\left(1-2{\cal H}_*\eta\right)^{1/2}~
\,\,\,for\,\,{\rm contracting}\,\,{\rm phase}, \label{leq} \nonumber\\
& & {a_* \over 1-{\cal H}_*\eta}~, \,\,\, for \,\,{\rm
inflationary}\,\,{\rm phase}, \label{geq} \ea respectively, where
${\cal H}_*$ sets the slow-roll inflationary scale by
$H_{inf}=H_*={\cal H}_*/a_*$.
Here, the pre-inflationary contraction is a kinetic-dominated
phase,
see Ref.\cite{Liu:2013kea} for a detailed model. In this mode, the
ghost-free bounce may be implemented in Einstein gravity, e.g.
\cite{Qiu:2011cy},\cite{Easson:2011zy},\cite{Osipov:2013ssa} and
\cite{Koehn:2013upa}. In addition, the bounce can also be realized
with modified gravity
\cite{Bamba:2013fha},\cite{Biswas:2005qr},\cite{Calcagni:2010bj}.
Actually, lots of the bounce mechanisms have been argued, see
\cite{Battefeld:2014uga},\cite{Lehners:2011kr} for reviews and
references. Generally the perturbation may continuously pass
through the bounce, and its spectrum is insensitive
 with respect to the implementing detail of the bounce, e.g.\cite{Battarra:2014tga}.
See also \cite{Liu:2010fm} for other case.

We, following Ref.\cite{Alexander:2004wk}, define the left-handed
and right-handed circular polarization modes $h_{s }$ with the
circular polarization tensor $p^{s}_{\ ij}$, and expand $h_{ij}$
as
\begin{equation}
{h}_{ij}(t, \mathbf{x}) = \sum\limits_{s = L,R}\int{d^3 \mathbf{k}
\over (2\pi)^3} h_{s }(t,\mathbf{k}) p^{(s)}_{
ij}e^{i\mathbf{k}\cdot\mathbf{x}}.
\end{equation}
where
$ik_q \epsilon^{rqj}p_{ij}^{(s)}=k \lambda_s p^{r(s)}_i$, and the
modes with $\lambda_{R,L} = 1,\,-1$ are called as the right-handed
mode and the left-handed mode, respectively.


Thus with (\ref{action1}), the equation of $h_{s}(t,\mathbf{k})$
is \be v_{sk}^{\prime\prime} +\left(k^2-{{z_s}^{\prime\prime}\over
z_s}\right) v_{sk} = 0, \label{uk2}\ee where $v_{sk} \equiv
z_s{h_{s}}$, and $ z_s=a \left(1 - \frac{\lambda_{s}
k}{a^2} \frac{f^\prime}{M_{P}^2}\right)^{1/2}$. Here, $f=\alpha
{\phi\over M_P}$ and $\alpha$ is a parameter determined by the
potential fundamental theory. Thus $z_s$ equals to
\be z_s=a\sqrt{1- \lambda_s  \Theta \left(-{k\over {\cal
H}}\right)},\ee where $\Theta ={\alpha H^2 \sqrt{2\epsilon} \over
M_P^2}$ and $\epsilon=-{{\dot H}/H^2}$. When $k^2\simeq
z_s^{\prime\prime}/z_s$, the perturbation mode is leaving the
horizon. When $k^2\ll z_s^{\prime\prime}/z_s$, the solution of
$h_s$ given by Eq.(\ref{uk2}) is \ba h_s & \sim &
C\,\,\,\,\, is\,\,\,{{\rm constant}}\,\,\,{ {\rm mode}}\label{C}\\
&or &\, D\int {d\eta\over z_s^2}\,\,\,\,\, is\,\,\,{{\rm
decaying}}\,\,\,{ {\rm mode}} . \label{D}\ea



\subsection{The unpolarized gravitational wave}


When the gCS term is negligible, which implies $z_s=a$, both $h_R$
and $h_L$ will be equal and $v_{Rk}=v_{Lk}=v_k$. We firstly
investigate this unpolarized case.

When $k^2\gg {a^{\prime\prime}\over a}$, i.e. the perturbation is
deeply inside the horizon, $v_k$ oscillates with a constant
amplitude, \be v_k\sim {1\over \sqrt{2k}} e^{-ik\eta}.
\label{ini}\ee When $k^2\ll {a^{\prime\prime}\over a}$, i.e. the
perturbation is far outside the horizon, in the pre-inflationary
contracting phase, the solution of Eq.(\ref{uk2}) is given by \ba
v_k=\sqrt{\frac{\pi}{4}(x+{k\over {\cal H}_*})}
H^{(1)}_{0}(x+{k\over {\cal H}_*}), \ea where $x={k/{\cal
H}_*}-k\eta$, $H^{(1)}_{0}$ is the 0th order Hankel function of
the first kind.
While in the slow-roll inflationary phase the solution of
Eq.(\ref{uk2}) is \ba & &{v_k} = x^{1/2}\left[C_1
H_{3/2}^{(1)}(x)+C_2 H_{3/2}^{(2)}(x)\right], \label{vki} \ea
where $H_{3/2}^{(1)}$ and $H_{3/2}^{(2)}$ are the 3/2th-order
Hankel function of the first kind and the second kind,
respectively, the parameters $C_1$ and $C_2$ are only dependent on
$k$.

The continuity of $h_{s}$ around the bounce gives $C_1$ and $C_2$.
Thus the GW spectrum is
\ba {\cal P}_{T} = \sum\limits_{s = L,R} {\cal P}_{T,s}=
 {\cal P}_{T,inf} {2k\over \pi {\cal H}_0}\left|C_1
-C_2\right|^2, \label{ps} \ea
where ${\cal P}_{T,s}={k^3\over \pi^2} \left|h_s /a\right|^2$,
 and ${\cal P}_{T,inf}={2{H}_{inf}^2\over \pi^2
M_P^2}$ is that of the slow-roll inflation. We plot ${\cal
P}_{T}(k)$ in Fig.1, ${\cal P}_{T}(k)$ is almost scale-invariant 
for $k>{\cal H}_*$, since the corresponding modes are
produced in slow-roll inflationary phase, while for $k<{\cal H}_*$
it gets a cutoff. The shape of ${\cal P}_{T}(k)$ is the same with
that of the scalar spectrum in Ref.\cite{Liu:2013kea}.

When $k\ll {\cal H}_*$, we have approximately \ba {\cal P}_{T}^{k<
{\cal H}_*} & \sim & (2+\ln\frac{4{\cal
H}_0}{k})^2\frac{k^3}{{\cal H}_0^3}{\cal P}_{T,inf}, \label{P1}\ea
which is the usual output of original Pre-big bang scenario
\cite{MGV},\cite{GV2}, i.e. $n_T\simeq 3$. While for $k \gg {\cal
H}_*$, we have approximately \be {\cal P}_{T}^{k > {\cal H}_*} \sim
(1+\frac{{\cal H}_*}{4k}\sin\frac{2k}{{\cal H}_*}){\cal
P}_{T,inf}, \ee which is almost scale-invariant, but with a
decaying oscillation. The result is consistent with the solid line
in Fig.1.


\subsection{The circularly polarized gravitational wave}

When the gCS term is not negligible, it will produce a difference
between the amplitudes of $h_R$ and $h_L$. Here, to quantify this
chirality, we define a chiral parameter $\Delta\chi$ as
${\cal P}_{T,s}=\left({1-\lambda_s\Delta\chi}\right){\cal P}_T/2$,
in which $-1\leq\Delta\chi \leq 1$ reflects the magnitude of
parity violation of primordial GW. This definition equals to that
in \cite{Saito:2007kt},\cite{Gluscevic:2010}, also
\cite{Satoh:2007gn} but with inverse sign.


When $k>{\cal H}_*$, the modes are produced in the slow-roll
inflationary phase. In slow-roll inflation, the power spectrum of
$h_s$ is, Ref.\cite{Alexander:2004wk}, \be {\cal P}^{k>{\cal
H}_*}_{T,s}\simeq {H_{inf}^2\over \pi^2 M_P^2}\left(1-\lambda_s
{\pi\Theta\over 2}\right), \label{pk1}\ee where the terms with
higher order $\Theta$ are neglected. With Eq.(\ref{pk1}), we have
\ba \Delta\chi_{inf} = {2{\cal P}_{T,L}^{k>{\cal H}_*}\over {\cal
P}_{T}^{k>{\cal H}_*}}-1 \simeq {\alpha \pi H_{inf}^2
\sqrt{2\epsilon_{inf}} \over M_P^2}. \label{infchi}\ea Here,
$H_{inf}^2\ll M_P^2$, which implies $\Delta\chi_{inf}$ is
negligible. However, in a stringy embedding, $\alpha\sim
\sqrt{g_{str}} {M_P^2\over
M_{10}^2}$, 
it has been argued in \cite{Alexander:2004wk} that for the
suitable values of the string scale $M_{10}$ and the string
coupling $g_{str}$, we might have $\alpha {H_{inf}^2 \over M_P^2}
\sim 1$. Thus Eq.(\ref{infchi}) may be written as \be
\Delta\chi_{inf}\lesssim \sqrt{\epsilon_{inf}}. \label{chiinf}\ee

We will estimate the chirality parameter $\Delta\chi_{pre-inf}$ in
the pre-inflationary contracting phase. For the inflation, the
constant mode (\ref{C}) is dominated, but for the contraction,
$D\int d\eta/z_s^2$ is dominated.

During the contraction, we have \ba h_s^{k<{\cal H}_*}  \sim \int
{d\eta \over a^2\left[1- \lambda_s  \Theta \left(-{k\over {\cal
H}}\right)\right]}\sim \int {d\eta \over a^2}\left[1+ \lambda_s
\Theta \left(-{k\over {\cal H}}\right)\right],\label{hs1}\ea where
$\Theta (-{k\over {\cal H}})\ll 1 $ is used since $\Theta<1$ and
the mode is far outside the horizon. Now, with (\ref{leq}) and
${\cal H}_*=a_*H_{inf}$,
we could obtain \ba \Delta \chi_{pre-inf} & = & {2{\cal
P}_{T,L}^{k<{\cal H}_*}\over {\cal P}_{T}^{k<{\cal
H}_*}}-1\nonumber\\ & \simeq &
2\alpha\sqrt{2\epsilon_{pre-inf}}{H_{inf}^2\over M_P^2}\left({k
\over {\cal H}_*}\right)\left(\int^{0-} { d\eta\over
a_*^2(1-2{\cal H}_*\eta)^3}\right)\left( \int^{0-} {d\eta \over
a_*^2(1-2{\cal H}_*\eta)}\right)^{-1} \nonumber\\ & \simeq &
\alpha \sqrt{2\epsilon_{pre-inf}} {{H}_{inf}^2\over
M_P^2}\left({k\over {\cal H}_*}\right). \label{ctrchi}\ea Thus
\be {{\Delta\chi_{pre-inf}}}\sim {\sqrt{\epsilon_{pre-inf}}\over
\sqrt{\epsilon_{inf}}} \left({k\over {\cal H}_*}\right)
\Delta\chi_{inf}, \label{chictr}\ee which implies that for $k\ll
{\cal H}_*$, ${\Delta\chi_{pre-inf}}$ is negligible, while around
cutoff scale, i.e. $k\sim {\cal H}_*$, ${\Delta\chi_{pre-inf}}$ is
far larger than $\Delta\chi_{inf}$.



\begin{figure}[t]
\begin{center}
\includegraphics[scale=0.6,width=7.0cm]{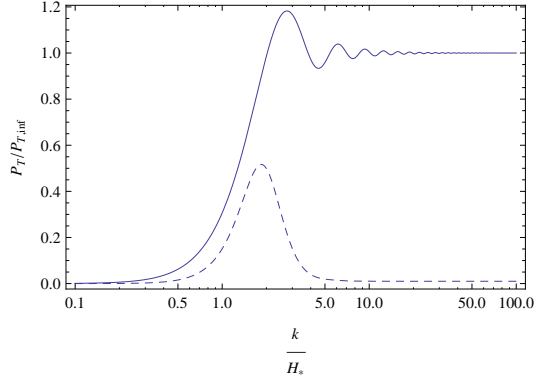}
\caption{In bounce inflation, ${\cal P}_T$ is plotted (solid
line), which for $k>{\cal H}_*$ is almost scale-invariant but with
a decaying oscillation and for $k<{\cal H}_*$ gets a cutoff. The
dashed line is $\Delta\chi{\cal P}_T$, in which $\Delta\chi$ is
the chirality parameter. \label{figure2:bestCMB0}}
\end{center}
\end{figure}

\begin{figure}
\begin{center}
\includegraphics[scale=0.6,width=8.0cm]{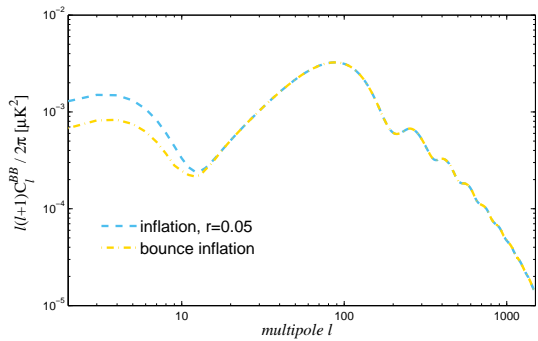}
\caption{Theoretical CMB BB-mode power spectrum for the slow-roll
inflation model (blue dashed line), in which $r=0.05$ is set for
our simulation, and the inflation model with the pre-inflationary
bounce (yellow dashed line). \label{figure2:bestCMB1}}
\end{center}
\end{figure}

In slow-roll inflationary scenario, $\Delta\chi$, which reflects
the parity violation of primordial GW, is scale invariant, and
also negligible. However, we see that if a pre-inflationary bounce
occurs, $\Delta\chi$ will show a bump around the
matching scale $k\sim {\cal H}_*$, see the dashed line in Fig.1
for a parameterized $\Delta\chi $ used in Sec.III. It could be
anticipated that this bump would leave the imprint in TB and
EB-mode power spectrum of CMB at large angular scale. Here, in the
leading order of $\Theta$, the polarized GW spectrum ${\cal P}_{T}
= \sum\limits_{s = L,R} {\cal P}_{T,s}$ is approximately
Eq.(\ref{ps}).

\section{CMB angular power spectra}

\subsection{BB-mode power spectrum}

In Ref.\cite{Liu:2013kea}, with the background (\ref{geq}), the
primordial scalar spectrum was calculated to fit the Planck+WMAP
data. The best-fit values of ${\cal H}_*$ and $A_{{\cal R}, inf}$
are $\ln \left({\cal H}_*/{\rm Mpc}^{-1}\right)=-8.60$ and $\ln
\left(10^{10} A_{{\cal R}, inf}\right)=3.084$.
With this result and Eq. (\ref{ps}), in which ${\cal P}_{T}^{inf}$
is parameterized as \be {\cal P}_{T,inf}=rA_{{\cal R},
inf}\left({k\over k_*}\right)^{n_{T,inf}},\ee
we plot the BB-mode power spectrum as Fig.2 by modifying the CAMB
\cite{camb}, in which $r=0.05$ is set for our simulation.
We see that the large-scale cutoff of the primordial GW spectrum
brings a BB-mode power suppression at $l<20$, which is a
significant prediction of pre-inflationary bounce.

\subsection{TB and EB-mode power spectra}

The TB or EB-mode power spectrum is \ba C_l^{T/E,B} \sim \int
{dk\over k } \,\,\Delta\chi\,{\cal P}_T \left[{\Delta_{l}^{
T/E}}(k){\Delta_{l}^{B}(k)}\right],
\ea which relies on the difference between left-handed $h_L$ and
right-handed $h_R$, i.e.$\Delta\chi {\cal P}_T$,
e.g.\cite{Saito:2007kt},\cite{Gluscevic:2010}. Here, $\Delta \chi$
may be phenomenologically parameterized as \ba |\Delta\chi |
 & = & \Delta\chi_{inf}+
 \nonumber\\ &  & ({\Delta\chi_{pre-inf}-
\Delta\chi_{inf}})\left[1/2-{{\rm Tanh}({1\over B}{\rm
Log}_{10}{k\over 2{\cal H}_*})\over 2}\right], \label{chip}\ea
where for $k>{\cal H}_*$, the chiral parameter $\Delta\chi$ equals
(\ref{infchi}) in slow-roll inflation, and for $k<{\cal H}_*$,
$\Delta\chi$ equals (\ref{ctrchi}) in the pre-inflationary
contraction. We plot $\Delta\chi {\cal P}_T$ in Fig.1, in which
since $\Delta\chi\sim \sqrt{\epsilon_{inf}}$ is negligible for
$k>{\cal H}_*$, while \be \Delta\chi\sim ({k\over {\cal H}_*})\ee
is large around $k\sim 2{\cal H}_*$ but is suppressed for $k\ll
{\cal H}_*$, a bump appears around $k\sim 2{\cal H}_*$.

Usually, the TB and EB-mode power spectrum should vanish. Here,
since the gravity is chiral, the amplitudes of left-handed mode
$h_L$ and right-handed mode $h_R$ of the primordial GWs are
different, which straightly induce non-vanishing TB and EB-mode
correlations at CMB last scatting surface.
We plot TB and EB-mode correlations in Fig.3. We can see that
compared with those in the slow-roll inflation model, the TB and
EB-mode spectra are enhanced on large scale due to the bump of
$\Delta\chi$ at corresponding scale.

The results can be explained as follows. The bump height of the TB
or EB-mode spectrum at $l<10$ is caused by the reionization of
universe, which is mainly depicted by the optical depth to the
beginning of reionization, $\tau$. The bump height of the BB and
TB-mode power spectra around $l\sim 2$ have been roughly estimated
in Ref.\cite{Saito:2007kt}, which are \be C_{l\sim 2}^{BB}\simeq
{1\over 100}\left(1-e^{-\tau}\right)^2 C_{T,l\sim 2}^{TT}, \ee \be
\left|C_{l\sim 2}^{TB}\right|\simeq {|\Delta\chi|\over
10}e^{-\tau}\left(1-e^{-\tau}\right) C_{T,l\sim 2}^{TT}, \ee where
$C_{T,l}^{TT}$ stands for the TT-mode power spectrum from the
primordial GW without reionization. Thus since ${\cal P}_T$ is cut
off on large scale, which lowers $C_{T,l\sim 2}^{TT}$, the
reionization bump in the BB-mode power spectrum is suppressed.
However, since the TB-mode power spectrum relies on
$|\Delta\chi|C_{T,l\sim 2}^{TT}$, if $\Delta\chi_{pre-inf}\gg
\Delta\chi_{inf}$, we may have an enhanced reionization bump,
compared with that in slow-roll inflation model.

\begin{figure}[t]
\begin{center}
\includegraphics[scale=0.6,width=9.0cm]{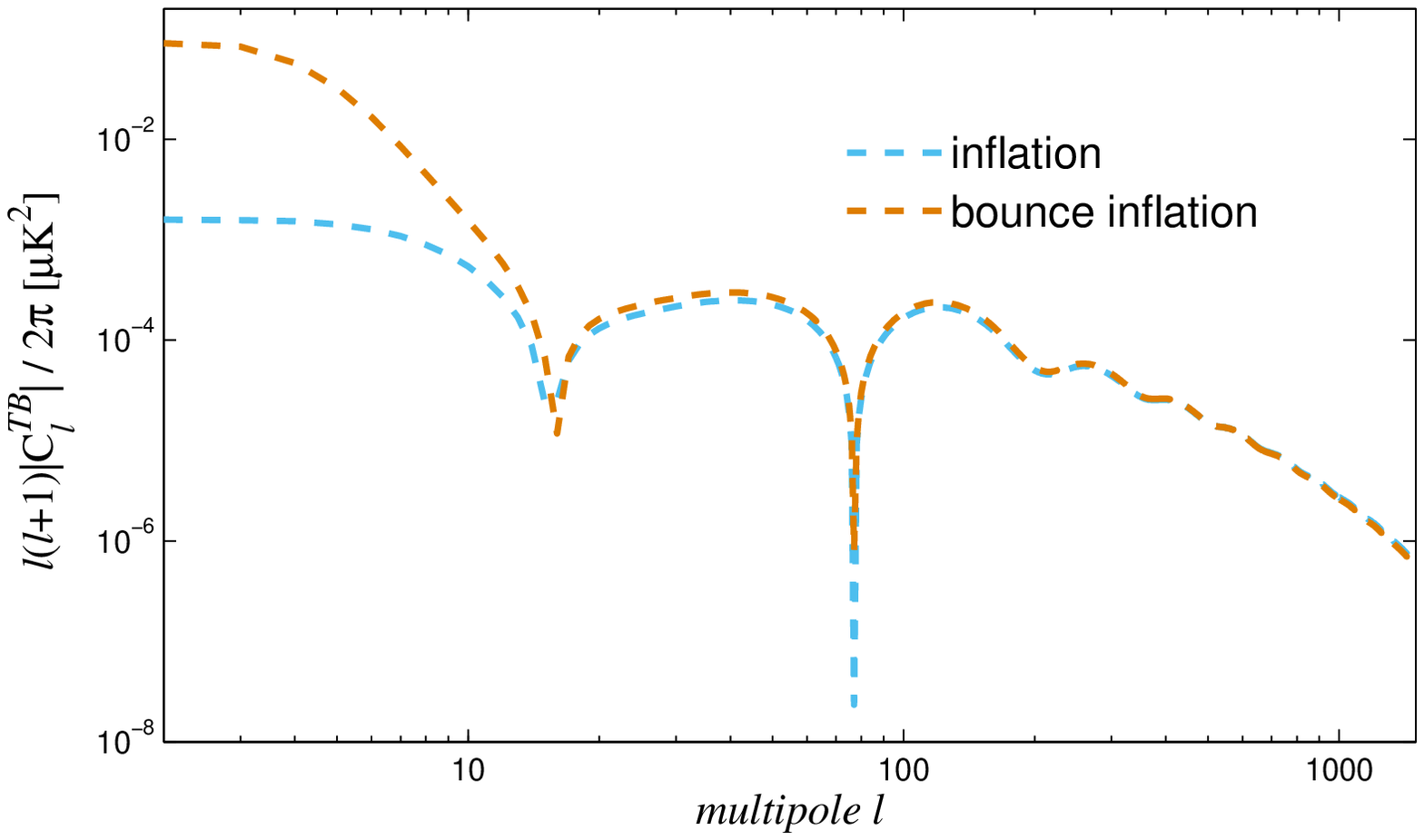}
\includegraphics[scale=0.6,width=9.0cm]{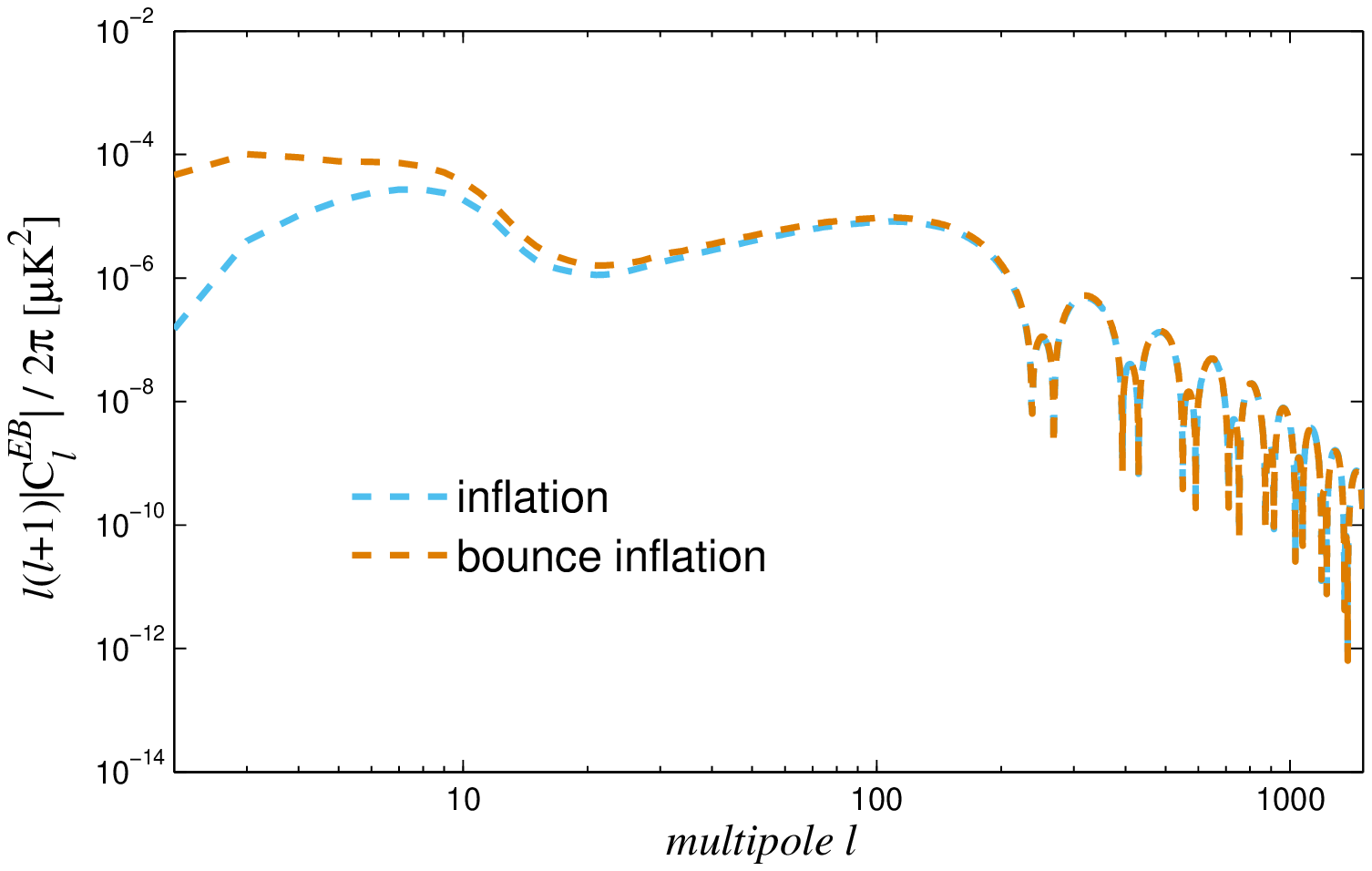}
\caption{Theoretical CMB TB and EB-mode power spectra for the slow-roll
inflation model (blue dashed line), in which $r=0.05$ and
$\Delta\chi_{inf}=0.01$, and the inflation model with the
pre-inflationary bounce (brown dashed line), in which
$\Delta\chi_{pre-inf}=0.5$. \label{figure2:bestCMB2}}
\end{center}
\end{figure}


However, it should be mentioned that for a different set of the
values of parameters in Eq.(\ref{chip}), which reflects the
details of different bounce mechanisms and different
pre-inflationary evolutions, the possibility that $|\Delta\chi|$
does not set off the effect of low $C_{T,l\sim 2}^{TT}$ can not
ruled out, though this case is quite fine-tuning. Thus, whether
the reionization bumps of the TB and EB-mode are enhanced or not might be
model-dependent. However, if their reionization bumps are
actually enhanced, this will be an interesting signature of the
pre-inflationary bounce, different from those in the TT-mode power
spectrum \cite{Liu:2013kea} and the BB-mode power
spectrum in Fig.2.

Finally, we discuss the feasibility to detect this signature of
pre-inflationary bounce.
Whether the effect of the parity violation is detectable or not is
dependent on the value of $r$. Ref.\cite{Gluscevic:2010} showed that if
$\Delta\chi<0.5$, the value of $r$ should be larger than $0.22$
for Planck and $r>0.03$ for the CMBPol, see also
\cite{Saito:2007kt},\cite{AQWT}. However, if one consider the
cosmic variance limit, it requires $r>0.008$. Thus the
determination of the chirality of primordial GW is quite difficult
if $r$ is not large enough. But if $\Delta\chi\sim\pm1$, the
detection will become much easier.
In the slow-roll inflation model without pre-inflationary
bounce, $\Delta\chi$ is negligible, thus it is hardly possible to
detect the signature of parity violation. However, in the model
with pre-inflationary bounce, before the inflation begins,
$\Delta\chi\sim 1$, which will makes the
signature of parity violation on corresponding scales become more detectable. It was found in
e.g.\cite{Gluscevic:2010} that if $|\Delta\chi|\sim 1$, we only
need $r>0.082$ for Planck, $r>0.0079$ for CMBPol and $r>0.0023$
for the cosmic variance limit. Thus the chirality of the primordial gravity might provide us an opportunity to test the
pre-inflationary physics.

\section{Discussion}

Recently, the Planck collaboration has shown a power deficit in
the CMB TT-mode power spectrum at low-$l$, which might imply the
occurrence of a pre-inflationary bounce, a solution to the initial
singularity problem of inflationary universe. In string landscape,
the bounce will induce a AdS-dS transition
\cite{Piao:2004me},\cite{Garriga:2013cix},\cite{Gupt:2013poa},
which is an efficient route to the slow-roll inflation, see the
Appendix in \cite{Liu:2013kea}, also see
\cite{Piao:2009ku},\cite{Liu:2014uda},\cite{Zhang:2010bb} for the
effect of perturbation.

The gCS term universally appears in particle physics and string
theory \cite{Witten83},\cite{Jackiw:2003pm}, which will lead to
the circular polarization of GW produced in primordial universe.
This circular polarization will induce TB and EB-mode correlations at
CMB last scatting surface,
which might provide an alternative test for the pre-inflationary
bounce.

We have calculated the circularly polarized GW, leaded by the gCS
term, in the inflation model with the pre-inflationary
bounce, and find that if the pre-inflationary bounce actually
occurs, the TB and EB-mode correlations on large scale
will be enhanced, while the BB-mode correlation on corresponding
scales is suppressed. This result applies to the cases that the
bounce can be realized without modified gravity,
e.g.\cite{Qiu:2011cy},\cite{Easson:2011zy},\cite{Osipov:2013ssa}
and \cite{Koehn:2013upa}, and with modified gravity but the parity
violation of gravity is mainly from the gravitational CS term.

However, it should be acknowledged that the amplitudes of both TB
and EB-mode correlations are determined by the value of $r$, thus for negligibly
small $r$ both TB and EB-mode power spectra are actually undetectable, even if
$\Delta\chi\simeq 1$. Thus rather than argue whether the detection
is possible or not, we would like to conclude that if the gravity
in primordial universe is chiral, or parity-violating, compared
with those produced during the slow-roll inflation, both TB and EB-mode correlations
around the pre-inflationary bounce may be overwhelmingly enhanced,
which is a significant character of the pre-inflationary bounce.

Here, the enhancement of both TB and EB-mode correlations on large scale actually
is a reflection of the background evolution of pre-inflationary
universe, since \be C^{T/E,B}_{l\sim 2}\sim \Delta\chi\sim
\sqrt{|\epsilon_{Pre-inf}|}.\ee Thus if the anomalies on large
scale in the TT-mode power spectrum is attributed to a pre-inflationary
evolution, it is interesting to show TB and EB-mode power spectra for other
non-slow-roll pre-inflationary background, and see whether the
result is similar, which will be investigated afterwards. In
addition, it is also interesting to consider other sources
inducing chiral GW, e.g. non-Abelian gauge fields
\cite{Noorbala:2012fh}.



\textbf{Acknowledgments}

We thank Mingzhe Li for helpful discussion and comment. This work
is supported by NSFC, No.11222546, and National Basic Research
Program of China, No.2010CB832804. We acknowledge the use of CAMB.

\end{document}